\documentclass[prd,showpacs]{revtex4}
\usepackage{epsfig}
\usepackage{amsmath}
\usepackage{amssymb}

\newcommand{\slashs}[1]{\not{\!#1}}
\begin{document}
\title{Rare decays of $B\to J/\psi D^{(*)}$ and $B\to \eta_c D^{(*)}$
in pQCD Approach}
\author{
 Ying Li\footnote{liying@mail.ihep.ac.cn}, Cai-Dian L\"u}
\affiliation{\it \small  CCAST (World Laboratory), P.O. Box 8730,
Beijing 100080, China;} \affiliation
 {\it \small Institute of High
Energy Physics, P.O.Box 918(4), Beijing 100049, China;}
\author{Cong-Feng Qiao\footnote{qiaocf@gucas.ac.cn}}
\affiliation{\it \small  CCAST (World Laboratory), P.O. Box 8730,
Beijing 100080, China;}
 \affiliation{\it\small  \it \small  Graduate School of the Chinese Academy of
Sciences, Beijing 100049, China }
\begin{abstract}
Motivated by the recent measurement of the upper limit of
${B}^0 \to J/\psi D$ branching ratio, which is important in
accounting for the soft $J/\psi$ production in $B$ decays,
we investigate ${B}^0 \to J/\psi D^{(\star)}$ and $\eta_c D^{(\star)}$
decays in perturbative QCD approach based on $k_T$ factorization.
Being pure annihilation (W-exchange) decays, these branching ratios
are estimated to be at the order of $10^{-5} \sim 10^{-7}$, which are
just at the corner of being observable at the $B$ factories. The measurements
of these decay channels may help us to understand the QCD dynamics in the
corresponding energy scale, especially the reliability of pQCD approach
to these processes.
\end{abstract}
\maketitle
\section{Introducrion}
In 1995, the CLEO Collaboration found a hump in the low momentum
region of the inclusive spectrum of $B \to J/\psi+X$ decay
\cite{Cleo}. Later on, this observation was confirmed by Belle
\cite{Belle} and BaBar \cite{Babar}. In these measurements, there
is an excess in the momentum spectrum of the $J/\psi$ recoiling
mass at $\sim 2~\mathrm{GeV}$. And, the excess corresponds to a
branching ratio of $6\times 10^{-4}$. In order to explain this
result, various hypotheses have been proposed \cite{brod, hou,
yang}.

In Ref. \cite{hou}, Chang and Hou employ the idea of intrinsic
charm \cite{Incharm} $c\bar c$ inside the $B$ meson to this issue.
Based on this scenario, they predicted that the branching ratio of
$B\to J/\psi D$ should be about $10^{-4}$. However, according to
recent BaBar and Belle measurements, the branching ratio upper
limit of this process is less than $10^{-5}$ \cite{ep1,ep2}, which
implies that the intrinsic charm mechanism is not favored. In
another scenario, in which the charmonium is produced
predominantly in the Color-Octet mechanism, Eilam and Yang
estimated the branching ratio of $B\to J/\psi D$ \cite{yang} and
got a result of about $10^{-8}$. However, in the collinear
factorization, they have to use a cut-off or $\delta$-function to
tame the end-point singularity. Hence, their numerical results are
not stable. The recent progress in perturbative QCD (pQCD)
treatment, based on the $k_T$ factorization, of $B$ meson decays
can solve this problem by introducing the Sudakov form factor
through the threshold resummation. Now, the pQCD approach
\cite{LY} has become one of the broadly used theoretical methods
in investigating the $B$ meson two-body non-leptonic decays. Base
on the pQCD approach, many $B$ meson decay modes have been
calculated, like $B \to K\pi, \pi\pi$ \cite{hep-ph/9411308}, etc.,
and most results are consistent with the experimental data. Since
there is no end-point singularity, the pQCD approach can also be
applied to the pure "annihilation processes ", such as $B \to D_s
K$ \cite{bdsk}.

In this work, we calculate the $B\to J/\psi D^{(*)}$ and $B\to
\eta_c D^{(*)}$ processes in the pQCD $k_T$ factorization. In the
decay of $B\to J/\psi D$, the $W$ boson exchange induces the four
quark operator $\bar c b \to \bar{u}d$, and an additional pair of
$c \bar{c}$ is created by a gluon. This gluon can attach to any
quark involving in the four-quark operator. In the rest frame of
$B$ meson, the produced $c$ and $\bar{c}$ quarks in the final
states have the momenta of order $\mathcal{O}(P_\psi/2)$ and
$\mathcal{O}(P_D/2)$, respectively. Therefore, the gluon, which
generates the charm quark pair, possesses a virtuality of order
$\sim \mathcal{O}(M_B/2)$, which enables the perturbative QCD
calculation reliable.

The paper is organized as follows: we present the formalism used
in the calculation of $B\to J/\psi D^{(*)}$ and $B\to \eta_c
D^{(*)}$ decays in Section \ref{3}. In Section \ref{4} we give out
the numerical calculation results and some discussion on them. The
last section is left for conclusions and summary.

\section{Kinematics}\label{3}

The effective Hamiltonian for decay modes $B\to J/\psi D^{(*)}$
and $B\to \eta_c D^{(*)}$ is given by \cite{Buchalla:1995vs}
\begin{eqnarray}
\label{heff} {\cal H}_{\rm eff} &=&\frac{G_{F}}{\sqrt{2}}  V_{cb}
V_{ud}^{*} \left[ C_{1}(\mu)O_2 +C_{2}(\mu)O_2\right],\\
\nonumber O_1 &=&\bar{c}\gamma_{\mu}(1-\gamma_{5})u\,
\bar{d}\gamma^{\mu}(1-\gamma_{5})b,\\
O_2 &=&\bar{d}\gamma_{\mu}(1-\gamma_{5})u\,
\bar{c}\gamma^{\mu}(1-\gamma_{5})b.
\end{eqnarray}
As usual, in the pQCD approach the momenta of the final states are
expressed in its light-cone components, like
\begin{eqnarray}
p=(p^+,
 p^-,\vec{p}_{T})=\left(\frac{p^0+p^3}{\sqrt{2}},\frac{p^0-p^3}{\sqrt{2}},(p^1,p^2)\right).
\end{eqnarray}
And, the decay amplitude can be generally written as:
\begin{eqnarray}
\mathcal{M} &\sim& \int\!\! dx_1 dx_2 dx_3 b_1db_1 b_2db_2
b_3db_3\  \nonumber \\
 &&\times \mathrm{Tr} \bigl[ C (t) \Phi_B (x_1,b_1)
\Phi_{\psi} (x_2,b_2) \Phi_{D} (x_3,b_3) H (x_i,b_i,t)e^{-S (t)}
\bigr]\; . \label{eq:convolution1}
\end{eqnarray}
Here, $\mathrm{Tr}$ denotes the trace over Dirac and color
indices. $C (t)$ is Wilson coefficient of the four quark operator
which results from the radiative corrections at short distance.
$\Phi_{M}$ denote the wave functions which are process independent
and represent the non-perturbative dynamics of hadronization. The
hard interaction kernel $H$ is, nevertheless, process-dependent
and can be calculated by perturbation QCD. $t$ is chosen as the
largest energy scale involving in the hard interaction to avoid
the largest logarithms.  $S (t)$ is Sudakov form factor resulted
from the resummation of double logarithms
\cite{hep-ph/9411308,hep-ph/9607214}. Therefore, in
eq.(\ref{eq:convolution1}) only the hard part is process dependent
and will be calculated in the following.

\subsection{The $B\to J/\psi D$ Decays}
Of the $B$- and $D^{(*)}$-meson wavefunctions, we make use of the
same parameterizations as used in the studies of different
processes \cite{hep-ph/9411308,0305335}. For vector $J/\psi$
meson, in terms of the notation in Ref.~\cite{TLS}, we decompose
the nonlocal matrix elements for the longitudinally and
transversely polarized $J/\psi$ mesons into
\begin{eqnarray}
\langle J/\psi(P,\epsilon_L)|\bar c(z)_jc(0)_l|0\rangle
&=&\frac{1}{\sqrt{2N_c}}\int_0^1 dx e^{ixP\cdot z}
\bigg\{m_{J/\psi}[\slashs \epsilon_L]_{lj}\Psi^L(x)+[\slashs
\epsilon_L\slashs P]_{lj} \Psi^{t}(x)
\bigg\}\;,
\label{lpf}\\
\langle J/\psi(P,\epsilon_T)|\bar c(z)_jc(0)_l|0\rangle
&=&\frac{1}{\sqrt{2N_c}}\int_0^1 dx e^{ixP\cdot z}
\bigg\{m_{J/\psi}[\slashs \epsilon_T]_{lj}\Psi^V(x)+
[\slashs\epsilon_T\slashs P]_{lj}\Psi^T(x)
\bigg\}\;, \label{spf}
\end{eqnarray}
respectively. Here, $\Psi^L$ and $\Psi^T$ denote for the twist-2
distribution amplitudes, and $\Psi^t$ and $\Psi^V$ for the twist-3
distribution amplitudes. $x$ represents the momentum fraction of
the charm quark inside the charmonium.

The $J/\psi$ meson asymptotic distribution amplitudes read as
\cite{BC04}
\begin{eqnarray}
\Psi^L(x)&=&\Psi^T(x)=9.58\frac{f_{J/\psi}}{2\sqrt{2N_c}}x(1-x)
\left[\frac{x(1-x)}{1-2.8x(1-x)}\right]^{0.7}\;,\nonumber\\
\Psi^t(x)&=&10.94\frac{f_{J/\psi}}{2\sqrt{2N_c}}(1-2x)^2
\left[\frac{x(1-x)}{1-2.8x(1-x)}\right]^{0.7}\;,\nonumber\\
\Psi^V(x)&=&1.67\frac{f_{J/\psi}}{2\sqrt{2N_c}}\left[1+(2x-1)^2\right]
\left[\frac{x(1-x)}{1-2.8x(1-x)}\right]^{0.7}\;,\label{jda}
\end{eqnarray}
in which the twist-3 ones $\Psi^{t,V}$ vanish, as the twist-2
ones, at the end points due to the factor $[x(1-x)]^{0.7}$. In
contrast to Ref.\cite{yang}, here we distinguish the longitudinal
and transverse distribution amplitudes of the polarized $J/\psi$,
which can exhibit the different asymptotic behaviors of these two
types.

\begin{figure}[htb]
\begin{center}
\includegraphics[scale=0.7]{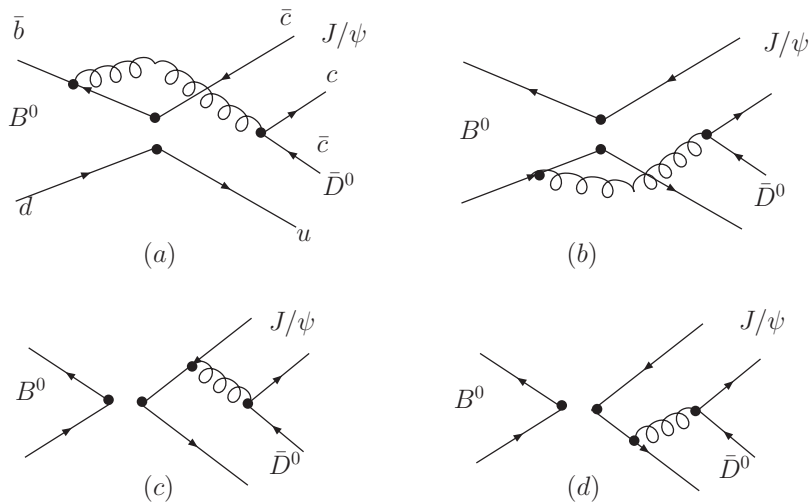}
\caption{Feynman diagrams for ${B}^0 \to J/\Psi \bar D^{0}$ decay
process in pQCD.} \label{figure:fig1}
\end{center}
\end{figure}

From the effective Hamiltonian (\ref{heff}), the Feynman diagrams
corresponding to the concerned process are drawn in Fig.1, where
the heavy dots denote the four quark operators. Similar figures
can be obtained by replacing the $J/\psi$ by $\eta_c$ for $B \to
\eta_c D$ process, and $D$ by $D^*$ for the vector $D$-meson
processes. With the meson wave functions and Sudakov factors, the
hard amplitude for factorizable annihilation diagrams Figs.1(a)
and (b) is
\begin{multline}
F_a = 16\pi C_F M_B^2 \int_0^1 \!\! dx_2 dx_3
 \int_0^\infty  \!\!  b_2 db_2\, b_3 db_3\ \phi_{D}(x_3) \\
\times \Bigl[  (x_3-1-x_3r_3^2-x_3 r_2^2) \Psi^L(x_2)
 E_{f}(t_1)
h_1(x_2,x_3,b_2,b_3) -\bigl\{[x_2-1+\\(1-2x_2)r_3^2+
(1-x_2)r_2^2]\Psi^L(x_2)+2x_2r_2r_3\Psi^t(x_2)
 \bigr\} E_{f}(t_2)
h_2(x_2,x_3,b_2,b_3) \Bigr].
\end{multline}
Here, the functions $E_f(t_a)$ contain Sudakov factors and Wilson
coefficients of four quark operator, and hard scale $t_a$. The
$h_a$, the virtual quark and gluon propagator, are given in the
appendix.

The result for the non-factorizable annihilation processes, shown
in Figs. 1(c) and (d), is
\begin{multline}
M_a  =  \frac{1}{\sqrt{2N_c}} 64\pi C_F M_B^2 \int_0^1 \!\! dx_1
dx_2 dx_3
 \int_0^\infty \!\! b_1 db_1\, b_2 db_2\
\phi_B(x_1,b_1) \phi_{D}(x_3) \\
\times \Bigl[ \bigl\{ (1-2r_2^2) \left(1-x_3\right)\Psi^L(x_2)
 + r_2 \left(x_2 -x_3\right) r_3 \Psi^t(x_2)
\bigr\}
E_{m}(t_3) h_3(x_1, x_2,x_3,b_1,b_2) \\
- \bigl\{
(1-x_2-(1-2x_2)r_2^2-(1-2x_2+x_3)r_3^2) \Psi^L(x_2) \\
  - r_2 (x_2-x_3) r_3 \Psi^t(x_2)
\bigr\} E_{m}(t_4) h_4(x_1, x_2,x_3,b_1,b_2) \Bigr].
\end{multline}
The total decay amplitude for this decay is:
\begin{equation}
{\cal A}_a (B\to J/\psi D)=f_BF_a+M_a.
\end{equation}
Thus, the $B$ meson decay width of the concerned process is:
\begin{equation}
 \Gamma(B\to J/\psi D) = \frac{G_F^2 M_B^3}{128\pi}(1-r_2^2-r_3^2)
\left| { V_{cb}V_{ud}^*\cal A}_a (B\to J/\psi D)\right|^2.
\end{equation}

\subsection{The $B\to \eta_c D^{(*)}$ Decays}
The nonlocal matrix element of $\eta_c$ production from vacuum can
be generally expressed as
\begin{eqnarray}
\langle \eta_{c}(P)|\bar c(z)_jc(0)_l|0\rangle
&=&\frac{1}{\sqrt{2N_c}}\int_0^1 dx e^{ixP\cdot z}
\bigg\{[\gamma_5\slashs P]_{lj}
\eta^{v}(x)+m_{\eta_{c}}[\gamma_5]_{lj}\eta^s(x) \bigg\}\; .
\label{epf0}
\end{eqnarray}
Here, $\eta^{v}(x)$ and $\eta^{s}(x)$ denote the twist-2 and
twist-3 $\eta_{c}$ meson distribution amplitudes, respectively.
The asymptotic forms of the $\eta_{c}$ distribution amplitudes are
given in \cite{BC04}:
\begin{eqnarray}
\eta^v(x)&=&9.58\frac{f_{\eta_{c}}}{2\sqrt{2N_c}}x(1-x)
\left[\frac{x(1-x)}{1-2.8x(1-x)}\right]^{0.7}\;,
\nonumber\\
\eta^s(x)&=&1.97\frac{f_{\eta_{c}}}{2\sqrt{2N_c}}
\left[\frac{x(1-x)}{1-2.8x(1-x)}\right]^{0.7}.
 \label{eda}
\end{eqnarray}

Performing the similar procedure as in above subsection, we can
get the decay amplitudes for $B\to \eta_{c} D$ and $B\to \eta_{c}
D^*$ straightforwardly.

\subsection{The $B\to J/\psi D^*$ Decays}
The $B\to J/\psi D^*$ decay rate are
\begin{equation}
\Gamma =\frac{G_{F}^{2}P_c}{32\pi M^{2}_{B} }
\sum_{\sigma=L,T}{\cal A}^{ (\sigma)\dagger }{\cal A^{
(\sigma)}}\;, \label{dr1}
\end{equation}
where $P_c\equiv |P_{2z}|=|P_{3z}|$ are the momenta of the
outgoing vector mesons; the superscript $\sigma$ denotes for the
helicity states of the two vector mesons, the $L$ for the
longitudinal and $T$ for the transverse components. The amplitude
$\cal M^{ (\sigma)}$ can be decomposed, according to the Lorentz
structure, to \cite{hep-ph/9810475}:
\begin{eqnarray}
{\cal A}^{ (\sigma)} &=&\epsilon_{2\mu}^{*}
(\sigma)\epsilon_{3\nu}^{*} (\sigma) \left[ a \,\, g^{\mu\nu} + {b
\over m_{\Psi}m_{D}} P_1^\mu P_1^\nu + i{c \over m_{\Psi}m_{D}}
\epsilon^{\mu\nu\alpha\beta} P_{2\alpha} P_{3\beta}\right]\;,
\nonumber \\
&\equiv &M_{B}^{2}{\cal A}_{L}+M_{B}^{2}{\cal A}_{N}
\epsilon^{*}_{2} (\sigma=T)\cdot\epsilon^{*}_{3} (\sigma=T)
+i{\cal A}_{T}\epsilon^{\alpha \beta\gamma \rho}
\epsilon^{*}_{2\alpha} (\sigma)\epsilon^{*}_{3\beta} (\sigma)
P_{2\gamma }P_{3\rho }\;,
\end{eqnarray}
with the convention $\epsilon^{0123} = 1$ for the total
anti-symmetric tensor and definitions
\begin{eqnarray}
M_B^2 \,\, {\cal A}_L &=& a \,\, \epsilon_2^{*} (L) \cdot
\epsilon_3^{*} (L) +{b \over m_{\Psi}m_{D}} \epsilon_{2}^{*} (L)
\cdot P_1 \,\, \epsilon_{3}^{*} (L) \cdot P_1\;,
\nonumber \\
M_B^2 \,\, {\cal A}_N &=& a \,\, \epsilon_2^{*} (T) \cdot
\epsilon_3^{*} (T)\;,
\label{id-rel} \\
{\cal A}_T &=& {c \over m_{\Psi}m_{D}}\;. \nonumber
\end{eqnarray}

Hereby, the only work left is to calculate the matrix elements
$A_L$, $A_N$ and $A_T$ with
\begin{eqnarray}
A_i=f_BF_i+M_i, (i=L,N,T).
\end{eqnarray}
Here, $F_i$ and $M_i$, coming from the calculation of hard
interaction, are given as follows:
\begin{multline}
F_L = 16\pi C_F M_B^2 \int_0^1 \!\! dx_2 dx_3
 \int_0^\infty  \!\!  b_2 db_2\, b_3 db_3\ \phi_{D}(x_3) \\
\times \Bigl[  (x_3-1-(x_3-2)r_3^2-(2x_3-1) r_2^2) \Psi^L(x_2)
 E_{f}(t_1)
h_1(x_2,x_3,b_2,b_3) \\+\bigl\{[x_2-1+(1-2x_2)r_3^2+
(1-x_2)r_2^2]\Psi^L(x_2)
 \bigr\} E_{f}(t_2)
h_2(x_2,x_3,b_2,b_3) \Bigr],
\end{multline}
\begin{multline}
F_N = 16\pi C_F M_B^2 \int_0^1 \!\! dx_2 dx_3r_2r_3
 \int_0^\infty  \!\!  b_2 db_2\, b_3 db_3\ \phi_{D}(x_3)r_2r_3 \\
\times \Bigl[  (1-x_3) \Psi^V(x_2)
 E_{f}(t_1)
h_1(x_2,x_3,b_2,b_3) +\bigl\{(x_2-2)\Psi^V(x_2)
 \bigr\} E_{f}(t_2)
h_2(x_2,x_3,b_2,b_3) \Bigr],
\end{multline}
\begin{multline}
F_T = 32\pi C_F M_B^2 \int_0^1 \!\! dx_2 dx_3r_2r_3
 \int_0^\infty  \!\!  b_2 db_2\, b_3 db_3\ \phi_{D}(x_3)r_2r_3 \\
\times \Bigl[  (1+x_3) \Psi^V(x_2)
 E_{f}(t_1)
h_1(x_2,x_3,b_2,b_3) +x_2\Psi^V(x_2)
 E_{f}(t_2)
h_2(x_2,x_3,b_2,b_3) \Bigr],
\end{multline}
\begin{multline}
M_L  =  \frac{1}{\sqrt{2N_c}} 64\pi C_F M_B^2 \int_0^1 \!\! dx_1
dx_2 dx_3
 \int_0^\infty \!\! b_1 db_1\, b_2 db_2\
\phi_B(x_1,b_1) \phi_{D}(x_3) \\
\times \Bigl[ \bigl\{
\left(x_3-1-2(x_3-1)(r_3^2+r_2^2)\right)\Psi^L(x_2)
 + r_2 \left(x_2 +x_3-2\right) r_3 \Psi^t(x_2)
\bigr\} \\
\times E_{m}(t_3) h_3(x_1, x_2,x_3,b_1,b_2) + \bigl\{
(1-x_2-(1-2x_2)r_2^2-(1-2x_2-x_3)r_3^2) \Psi^L(x_2) \\
  - r_2 (x_2+x_3) r_3 \Psi^t(x_2)
\bigr\} E_{m}(t_4) h_4(x_1, x_2,x_3,b_1,b_2) \Bigr],
\end{multline}
\begin{multline}
M_N  =  \frac{1}{\sqrt{2N_c}} 64\pi C_F M_B^2 \int_0^1 \!\! dx_1
dx_2 dx_3
 \int_0^\infty \!\! b_1 db_1\, b_2 db_2\
\phi_B(x_1,b_1) \phi_{D}(x_3) \\
\times \Bigl[ \bigl\{  \left(x_3-1\right)r_3^2\Psi^T(x_2)
 + \left(x_2 -1\right) r_2^2 \Psi^T(x_2)
\bigr\}
E_{m}(t_3) h_3(x_1, x_2,x_3,b_1,b_2) \\
+ \bigl\{ (-x_3r_3^2
\Psi^T(x_2)+2r_2r_3\Psi^V(x_2)-x_2r^2_2\Psi^T(x_2)
 \bigr\} E_{m}(t_4) h_4(x_1, x_2,x_3,b_1,b_2) \Bigr],
\end{multline}
\begin{multline}
M_T  =  \frac{1}{\sqrt{2N_c}} 128\pi C_F M_B^2 \int_0^1 \!\! dx_1
dx_2 dx_3
 \int_0^\infty \!\! b_1 db_1\, b_2 db_2\
\phi_B(x_1,b_1) \phi_{D}(x_3) \\
\times \Bigl[ \bigl\{  \left(x_3-1\right)r_3^2\Psi^T(x_2) -
\left(x_2 -1\right) r_2^2 \Psi^T(x_2) \bigr\}
E_{m}(t_3) h_3(x_1, x_2,x_3,b_1,b_2) \\
+ \bigl\{ (-x_3r_3^2 \Psi^T(x_2)+x_2r^2_2\Psi^T(x_2)
 \bigr\} E_{m}(t_4) h_4(x_1, x_2,x_3,b_1,b_2) \Bigr].
\end{multline}
\section{Numerical Results}\label{4}
In this work, the input parameters for the numerical calculation
are \cite{pdg}, which are commonly used in literature,
\begin{eqnarray}
m_{J/\psi}= 3.097 ~\mathrm{GeV}, m_{\eta_c}=2.980~\mathrm{GeV},
f_{D^*}=230 ~\mathrm{MeV},\nonumber\\ f_D=240 ~\mathrm{MeV},
m_D=1.87 ~\mathrm{GeV},m_{D^*}=2.005 ~\mathrm{GeV},\nonumber\\
m_B= 5.28 ~\mathrm{GeV}, |V_{cb}|=0.043,
|V_{ud}|=0.975,\tau_B=1.54\times 10^{-12}~\mathrm{s}.
\end{eqnarray}

At leading order, the main uncertainty comes from the meson wave
functions. Fortunately, the meson wave function, that describes
hadronic process, is universal at a certain scale. For instance,
the $B$ meson wave function is constrained by the measured
exclusive hadronic decays, like $B \to \pi\pi, K
\pi$\cite{hep-ph/9411308} with parameter $\omega_B$ from $0.32$ to
$0.48$. To determine the $D$ meson wave function is more tough
task than that of $B$ meson, because the heavy quark limit here is
not as good as in the $B$ meson case. Referring to to $B \to
D^{(*)}M$\cite{0305335} process, we can fit the $D$ meson wave
function parameter to be $a_D=0.8\pm 0.2$. The charmonium
distribution amplitudes can be inferred from the non-relativistic
heavy quarkonium bound state wave functions, which have been shown
to be successful in describing the charmonium production in
$e^+e^-$ collisions\cite{BC04}. The meson decay constant can be
measured via its pure leptonic decay. We have $f_{J/\psi}= 405\pm
14 ~\mathrm{MeV}$ and $f_{\eta_c}=420\pm 50 ~\mathrm{MeV}$. In
addition to the uncertainties remaining in the above input
parameters, the higher order corrections to the hard part are also
important, which is discussed in Ref. \cite{hsiangnan}.

Considering of the above uncertainties discussed, we can give out
the branching ratios of the discussed processes with error bars:
\begin{eqnarray}
{\rm Br}({B}^{0}\to J/\psi D)&=&(3.45^{+1.22} _{-1.46} \pm 1.51 \pm 0.32)\times10^{-6}, \nonumber \\
{\rm Br}({B}^{0}\to \eta_c D)&=&(1.28^{+0.32} _{-0.41}\pm 0.58 \pm 0.35)\times10^{-5}, \nonumber \\
{\rm Br}({B}^{0}\to \eta_c D^*)&=&(8.26^{+2.82} _{-2.34}\pm 2.23 \pm 2.06)\times10^{-6}, \nonumber \\
{\rm Br}({B}^{0}\to J/\psi D^*)&=&(7.04^{+2.43} _{-2.54}\pm 2.72
\pm 0.53)\times10^{-7}.
\end{eqnarray}
In the above, the uncertainties mainly come from $\omega_B$,
$a_D$, and the decay constants, respectively. To diminish the
uncertainties, for $B^{0}\to J/\psi D^*$ process, we evaluate the
longitudinal polarization fraction, that is:
\begin{equation}
P_{L}=\frac{\Gamma_{L}}{\Gamma}=0.66.
\end{equation}
This polarization fraction is not sensitive to the above mentioned
input parameters, because they only give an equally change of each
polarization amplitudes. However, this fraction is still sensitive
to the $J/\psi$ wave function. If we set the distribution
amplitude of transversal part the same as longitudinal part, the
branching ratio become larger and the polarization fractions
changed:
\begin{eqnarray}
&{\rm Br}({B}^{0}\to J/\psi D^*)=10.5 \times 10^{-7}.\\
&P_{L}=\frac{\Gamma_{L}}{\Gamma}=0.40;~~~
P_{N,T}=\frac{\Gamma_{N,T}}{\Gamma}=0.30.
\end{eqnarray}
That is to say that for ${B}^{0}\to J/\psi D^*$ the most important
uncertainty comes from the vector meson wave functions.

Compared to Ref. \cite{yang}, our results are much bigger. In
\cite{yang}, all wave functions, which describe the
non-perturbative hadronization, are $\delta$-function-like.
However, the $\delta$-like wave function can not embody the
relativistic corrections, though it can be used to avoid the
end-point singularity due to the wave function overlap absent. In
this work, the hadron distribution amplitudes are obtained from
from the established models with experimental fittings. In our
work, we take into account the Sudakov form factor and the
transverse momentum $k_T$ distribution, which are unique
characters of pQCD approach. For $B^{0}\to J/\psi D^*$ process,
since the charmonium longitudinal distribution amplitude is
different from its transverse one, and hence our longitudinal
polarization fraction are larger than what obtained in Ref.
\cite{yang}.

Since there is only one kind of CKM phase involving in the
concerned process, there should be no CP violation in these
process within the standard model. On experimental side, so far
there is only an upper limit for the branching ratio of
${B}^{0}\to J/\psi D$ process. That is
\begin{eqnarray}
{\rm Br}({B}^{0}\to J/\psi D)&<&1.3   \times 10^{-5}~~  [8], \nonumber \\
{\rm Br}({B}^{0}\to J/\psi D)&<&2.0   \times 10^{-5}~~[9],
\end{eqnarray}
from different experiment group, which is larger than, but very
close to our prediction.
\section{Summary}\label{5}
In this work, we have calculated the decays of ${B}^{0}\to J/\psi
(\eta_c) D^{(*)}$ in the pQCD approach. These $B$ meson exclusive
decay processes are in pure annihilation type, which is hard to be
accurately calculated in other approaches with the end-point
singularity. By keeping the transverse momentum $k_T$, the
end-point singularity disappears in our calculation. Our numerical
results shows that the branching ratios of ${B}^0\to \eta_c D$,
${B}^0\to \eta_c D^*$, ${B}^0\to J/\psi D$ and ${B}^0\to J/\psi
D^{(*)}$ decay processes are of the order $10^{-5}$, $10^{-6}$,
$10^{-6}$, and $10^{-7}$, respectively, which is just close to the
experiment capability to measure them. Although both Belle and
BaBar measured the $J/\psi$ momentum spectrum in $B$ inclusive
decays, they did not obtain the branching ratios of these
exclusive decays modes. Considering that the upper limits set by
experiments are very close to our predictions. We suggest that
BaBar and Belle measure these exclusive processes in near future.
The observation of these exclusive processes may greatly improve
our understanding on the $B$ meson exclusive hadronic decays, and
the corresponding theory describing them as well.
\section*{Acknowledgments}

This work was partly supported by the National Science Foundation
of China. Y. Li thanks J.-X Chen, Y.-L Shen, W. Wang, X.-Q Yu and
J. Zhu for useful discussions.

\begin{appendix}

\section{Some functions}
The function $E_f^i$, $E_m$, and $E'_m$ including Wilson
coefficients are defined as
\begin{gather}
 E_{f}(t) = ( C_1(t)+\frac{C_2(t)}{N_c} ) \alpha_s(t)\, e^{-S_\Psi(t)-S_D(t)}, \\
 E_{m}(t) = C_2(t) \alpha_s(t)\, e^{-S_B(t)-S_\Psi(t)-S_D(t)}.
\end{gather}
 where $S_B$, $S_\Psi$, and $S_D$ result from summing both double logarithms
caused by soft gluon corrections and single ones due to
the renormalization of ultra-violet divergence.
The above $S_{B,\Psi, D}$ are defined as
\begin{gather}
S_B(t) = s(x_1P_1^+,b_1) +
2 \int_{1/b_1}^t \frac{d\mu'}{\mu'} \gamma_q(\mu'), \\
S_\Psi(t) = s(x_2P_2^+,b_3) +
2 \int_{1/b_2}^t \frac{d\mu'}{\mu'} \gamma_q(\mu'), \\
S_D(t) = s(x_3P_3^-,b_3) +
2 \int_{1/b_3}^t \frac{d\mu'}{\mu'} \gamma_q(\mu'),
\end{gather}
where $s(Q,b)$, so-called Sudakov factor, is given in
Reference\cite{Li:1999kn}.

The functions $h_{i=1,2,3,4}$  in the decay
amplitudes come from the propagator of virtual quark and gluon.
They are defined by
\begin{align}
& h_1(x_2,x_3,b_2,b_3) = \left( \frac{\pi i}{2}\right)^2
H_0^{(1)}(M_B\sqrt{(x_2-1)(x_2-x_3)r_2^2-(x_3-1)(x_2r_3^2-x_2+1)}\, b_2) \nonumber \\
&\times \left\{
H_0^{(1)}(M_B\sqrt{1-x_3+x_3r_2^2-r_3^2}\, b_2)
J_0(M_B\sqrt{1-x_3+x_3r_2^2-r_3^2}\, b_3)
\theta(b_2 - b_3) + (b_2 \leftrightarrow b_3 ) \right\},
\\
& h_2(x_2,x_3,b_2,b_3) = \left( \frac{\pi i}{2}\right)^2
H_0^{(1)}(M_B\sqrt{(x_2-1)(x_2-x_3)r_2^2-(x_3-1)(x_2r_3^2-x_2+1)}\, b_2) \nonumber \\
&\times \left\{
H_0^{(1)}(M_B\sqrt{1-x_2+x_2r_3^2-x_2r_2^2}\, b_2)
J_0(M_B\sqrt{1-x_2+x_2r_3^2-x_2r_2^2}\, b_3)
\theta(b_2 - b_3) + (b_2 \leftrightarrow b_3 ) \right\},
\\
& h_{(j=3,4)}(x_1,x_2,x_3,b_1,b_2) = \nonumber \\
& \biggl\{
\frac{\pi i}{2} \mathrm{H}_0^{(1)}(M_B\sqrt{(x_2-1)(x_2-x_3)r_2^2-(x_3-1)(x_2r_3^2-x_2+1)}\, b_1)\nonumber \\
& \times \mathrm{J}_0(M_B\sqrt{(x_2-1)(x_2-x_3)r_2^2-(x_3-1)(x_2r_3^2-x_2+1)}\, b_2) \theta(b_1-b_2)
\nonumber \\
& \qquad\qquad\qquad\qquad + (b_1 \leftrightarrow b_2) \biggr\}
 \times\left(
\begin{matrix}
 \mathrm{K}_0(M_B F_{(j)} b_1), & \text{for}\quad F^2_{(j)}>0 \\
 \frac{\pi i}{2} \mathrm{H}_0^{(1)}(M_B\sqrt{|F^2_{(j)}|}\ b_1), &
 \text{for}\quad F^2_{(j)}<0
\end{matrix}\right),
\label{eq:propagator2}
\end{align}
where $\mathrm{H}_0^{(1)}(z) = \mathrm{J}_0(z) + i\, \mathrm{Y}_0(z)$, and
$F_{(j)}$s are defined by
\begin{eqnarray}
F^2_{(3)} &=& (x_1+x_2-1)(x_2-x_3)r_2^2+(x_3-1)(x_1+x_2-1-x_2r_3^2), \nonumber \\
F^2_{(4)} &=& -(x_1
-x_2)(x_2-x_3)r_2^2-x_2x_3r_3^2-x_1x_3+x_2x_3-1.
\end{eqnarray}

The hard scale $t$'s in the amplitudes are taken as the largest energy
scale in the $H$ to kill the large logarithmic radiative corrections:
\begin{gather}
 t_1 = \mathrm{max}(M_B \sqrt{1-x_3+x_3r_2^2-r_3^2},1/b_2,1/b_3), \\
 t_2 = \mathrm{max}(M_B \sqrt{1-x_2+x_2r_3^2-x_2r_2^2},1/b_2,1/b_3), \\
 t_j = \mathrm{max}(M_B \sqrt{|F^2_{(j)}|},
M_B \sqrt{(x_2-1)(x_2-x_3)r_2^2-(x_3-1)(x_2r_3^2-x_2+1)},
1/b_1,1/b_2).
\end{gather}

\end{appendix}

\end{document}